# PolySilicate Porous Organic Polymers (PSiPOPs), a new family of porous, ordered 3D reticular materials with polysilicate nodes and organic linkers

*Jelle Jamoul[a], Sam Smet[a], Sambhu Radhakrishnan[a,b], C. Vinod Chandran[a,b], Johan A. Martens[a,b],*

*Eric Breynaert[a,b]\**

[a] Centre for Surface Chemistry and Catalysis – Characterization and Application Team (COK-KAT), KU Leuven, Celestijnenlaan 200F Box 2461, 3001-Heverlee, Belgium

[b] NMRCoRe - NMR/X-Ray platform for Convergence Research, KU Leuven, Celestijnenlaan 200F Box 2461, 3001-Heverlee, Belgium

\* Email: Eric.Breynaert@kuleuven.be

**Abstract**

Spherosilicate, consisting of a double 4-ring cyclosilicate core (D4R; $Si_8O_{20}$) with every corner functionalized with a dimethylsilyl chloride group (-$SiMe_2Cl$), was used as node to construct an iso-reticular series of porous expanded network materials. Interconnecting the nodes with linear, aliphatic α,ω-alkanediol linker molecules yields PolySilicate Porous Organic Polymers (PSiPOPs), a new type of ordered reticular material related to the well-known metal-organic and covalent organic frameworks (MOFs & COFs). In the synthesis, sacrificial hydrogen-bonded $Si_8O_{20}$ cyclosilicate crystals are first converted into silyl chloride terminated spherosilicate. In a second step, these nodes are linked up by alkanediol units via the intermediate formation of a Si-N bond with catalytic amines such as pyridine and dimethylformamide. Overall, the presented synthesis converts D4R cyclosilicate into an ordered reticular framework with $[Si_8O_{20}]$-$[Si(CH_3)_2$-$]_8$ nodes and O-$(CH_2)_n$-O linkers. Example materials with ethylene glycol, 1,5-pentanediol, and 1,7-heptanediol as linker (n = 2, 5, and 7) were produced and characterized. On a macroscopic level, the synthesis yields porous frameworks exhibiting a thermal stability up to 400°C and a chemical stability between pH 1 and 12. $N_2$ physisorption revealed a secondary mesopore structure, indicating future options to produce hierarchical materials using soft templates. The molecular level structure of these reticular PSiPOP materials was elucidated using an NMR crystallography approach implementing a combination of 1D and 2D $^1H$ and $^{29}Si$ solid-state MAS NMR spectroscopy experiments. Previously reported reticular COF/POP materials implementing D4R-based nodes, used $Si_8$ octakis (phenyl) D4R POSS as a node, connecting it to the linker via a Si-C bond instead of a Si-O-C linkage.

**Introduction**

Reticular synthesis, interconnecting rigid metal oxide or lightweight non-metal element nodes with covalently bound, flexible, organic linker molecules, has yielded multiple families of crystalline porous materials (Figure 1),[1] including: metal-organic frameworks (MOFs)[2–6], covalent organic frameworks (COFs),[7–9] and crystalline covalent triazine frameworks (CTFs) (Figure 2a).[10,11] Porous organic polymers (POPs) can be considered the amorphous variant of COFs and CTFs, exhibiting a similar composition, porosity, and reticular structure but lacking crystalline order.[12,13] Applications of these families of reticular materials currently include gas storage and separation, catalysis, chemical sensing, biomedical applications, and electronic and ionic conduction.[14]

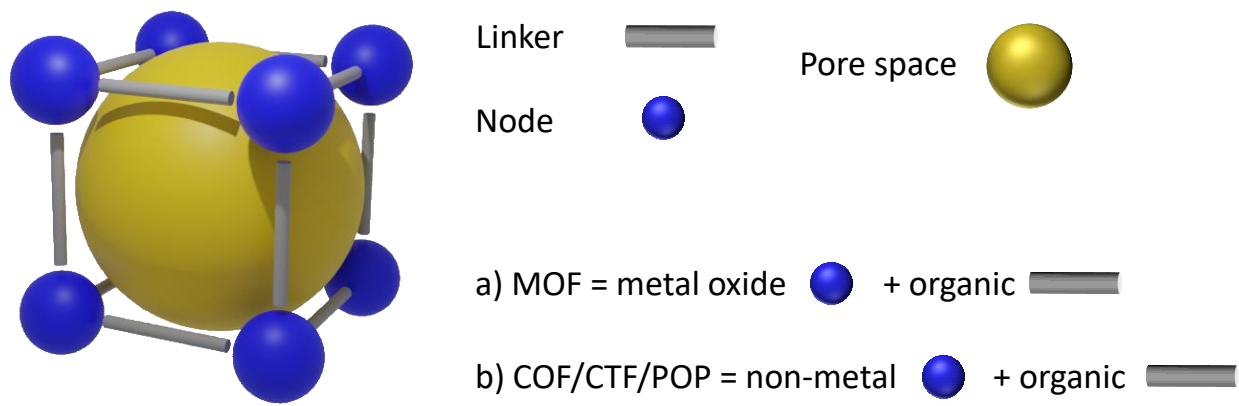

**Figure 1.** Molecular components and pore space in reticular porous materials such as MOFs, COFs, CTFs, and POPs.

$Zn_4O(BDC)_3$ (BDC = 1,4-benzendicarboxylate) was the first reported MOF with permanent porosity, exhibiting a specific surface area of ca. 300 m²/g and a pore volume of ca. 0.09 cm³/g.[2,15] In the following two decades, the MOF family has exponentially expanded with over 100.000 MOFs currently reported.[4,14,16–18] Synthesising MOFs with the same topology, simply by varying the length and/or side chain functionality of the organic linkers was coined isoreticular synthesis and produced for example the IRMOF series.[3,19] This strategy allows tailoring pore size and volume, as well as chemical functionality of the pore walls while the node remains invariant.[3,19–22] Initial COF materials contained boroxine nodes, linked into 2D layers with aromatic molecules.[7]

Ordered π- π stacking of COF layers resulted in mostly uniform one-dimensional pores in the micro- to small mesopore range.[7,8] Today, also aldehydes, amines, catechols, nitriles, hydrazine/hydrazides, and heteroatoms like phosphor have been incorporated into COF materials.[8–10,23–27] CTFs, a subgroup of the COF family, are built up from triazine nodes, i.e. aromatic C=N rings, covalently interlinked by organic molecules.[10,11] While MOFs and zeolites are typically 3D porous frameworks, the majority of COFs and CTFs are 2D layered frameworks.[8] Recently, also silicon - the base element of alumino-silicate zeolites - has found its way to porous 3D reticular organic materials, initially as a support for COF – silica hybrid materials consisting of COFs grafted on amino-substituted silica surfaces (amorphous gels and silicate double-four ring polyhedral oligomeric silsesquioxanes, i.e. D4R POSS), thus resulting in a reversed-phase silica with core-shell geometry.[28–33] More recently silicate also has been implemented in reticular materials as a node. Using a similar strategy as for MOFs, COF-silicate hybrid materials, i.e. SiCOFs, were synthesized with ionic, single-atom silicon centres ($SiO_6^{2-}$ or $SiF_6^{2-}$ octahedra, and $SiO_4R^-$) as nodes.[34–37] Related amorphous materials have been designated SiPOPs.[38–41] Asides inorganic monomeric silicon centres, also monomeric organosilicon centres ($SiR_4$) have been used as node in hybrid structures resembling MOFs. Exploiting chelating functionalities on the R-groups of the $SiR_4$ node, organic linkers were linked to the node through a metal ion coordinating both the $SiR_4$ node and the linker.[42] Very few COF materials containing oligomeric Si-containing nodes have been reported until now. A first example exploits a trigonal bipyramidal borosilicate core (($BR)_3(SiR')_2O_6$, 2 Si atoms) as node for a COF material,[43] but there also exist COF materials (COF-S7 and COF-S12) with an organosilicate D4R POSS ($Si_8O_{12}R_8$, 8 Si atoms) node. In the synthesis of both COF materials, octakis (phenyl) POSS is converted, via a nitro intermediate, to aminophenyl POSS which serves as the node for the final COF materials.[44] In general, such POSS and related spherosilicate materials are synthesized via hydrolysis and condensation reactions starting from trichloro or trialkoxy alkylsilanes, usually yielding a majority of random polymeric structures due to unselective conditions. The tedious and lengthy synthesis of such nodes impacts the yield of the POSS or spherosilicate node and of the final COF materials.[44–48] Related amorphous materials based on (mostly $Si_8$) silicate D4R nodes and organic linkers (mostly aromatics but also diols) were reported but contain nodes with a lower Si nuclearity and exhibit the disadvantage of non-selective node synthesis.[49–54]

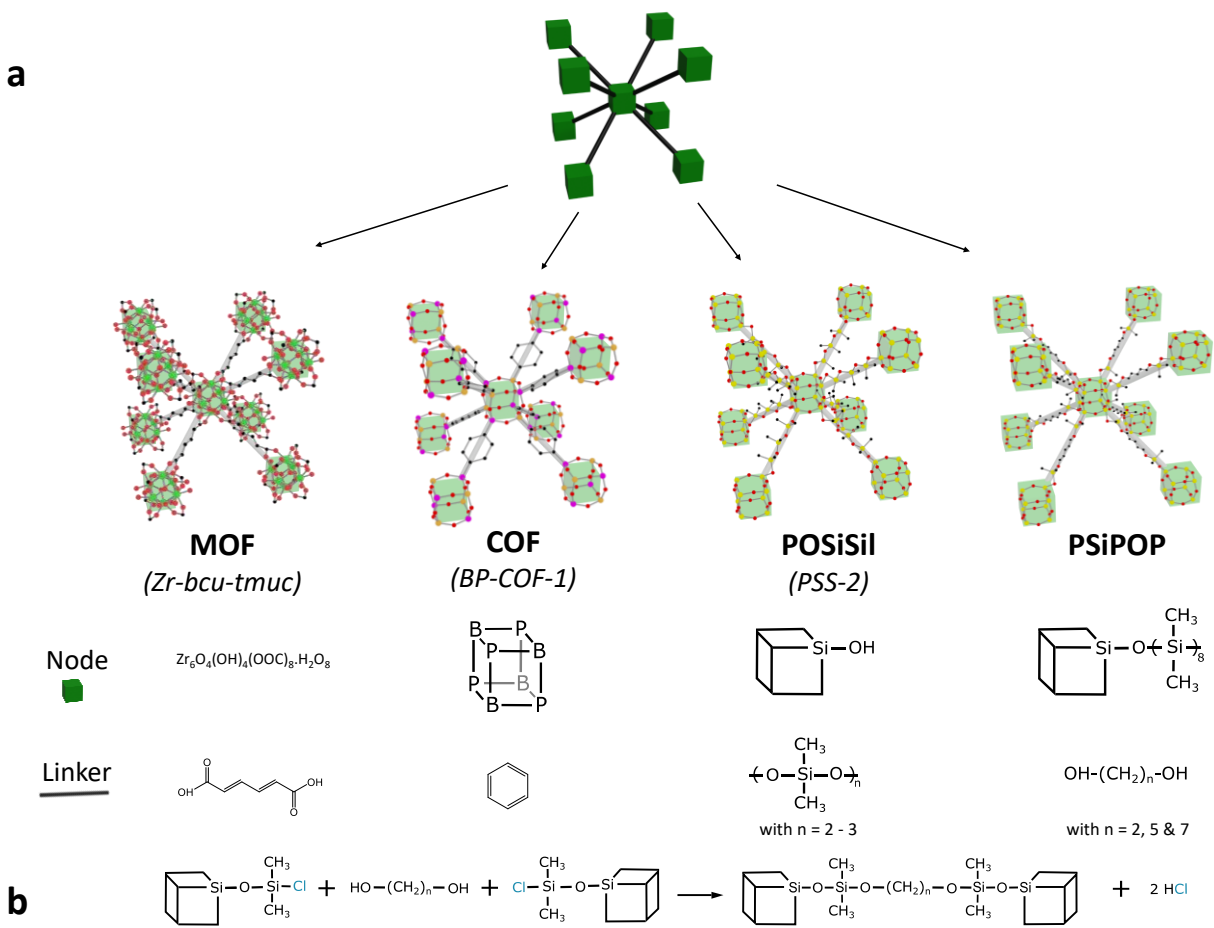

**Figure 2.** a. Comparison of reticular materials with similar node-linker construction: MOFs, COFs, POSiSils & PSiPOPs. The MOF with bcu topology uses a hexanuclear Zr oxide node interconnected with trans,trans-muconic acid (tmuc).[18] An 8-folded boron-phosphorus cubic node, with alternating P and B atoms on the corners and oxygen in between, and phenyl linkers resulted in BP-COF-1.[27] PSS-2 is constituted using silicate D4R node ($Si_8O_{20}$) and dimethylsilicone linkers.[56] Interconnecting octakis (dimethylsilyl chloride) spherosilicates using linear, aliphatic α,ω-alkanediols (ethylene glycol, 1,5-pentanediol, and 1,7-heptanediol) yields PSiPOPs; new reticular materials exploiting the chemistry of Si-Cl bonds and alcohol functional groups. Colour code: black, C; red, O; yellow, Si; blue, N; green, Zr; purple, P; orange, B. Protons are omitted for clarity. b. Schematic representation of coupling dimethylsilyl chloride spherosilicates with linear, aliphatic α,ω-alkanediols. For clarity cubes show only one functionalized corner.

Our group recently demonstrated a highly efficient synthesis of octakis (dimethylsilyl chloride) spherosilicate ($[Si_8O_{12}][OSi(CH_3)_2Cl]_8$; 16 Si atoms) and implemented this unit as a node in the reticular synthesis of POSiSils, 3D amorphous silicate-silicone hybrid polymers (Figure 2a).[55,56] Key novelty in this last synthesis was the use of sacrificial, D4R-based, hydrogen-bonded cyclosilicate hydrate (CySH) crystals, templated by tetrabutylammonium in an LTA-type topology. Reacting these crystals with dimethyldichlorosilane yielded a pure source of $[Si_8O_{12}][OSi(CH_3)_2Cl]_8$ spherosilicates, which were then hydrolyzed to generate PSS-2.[55–57]

Building on the innovation of Smet et al.,[55] the here reported work exploits the potential of $[Si_8O_{12}][OSi(CH_3)_2Cl]_8$ spherosilicates with reactive Si-Cl bonds on each corner, as source of nodes for the synthesis of a new family of porous, 3D reticular hybrid materials. Reacting the terminal Si-Cl bonds of dimethylsilyl chloride spherosilicates with linear, aliphatic α,ω-alkanediols (Figure 2b) readily generates a covalently bound, porous, 3D reticular organic polymer consisting of silicate/silicone nodes, interconnected by organic linkers. In analogy of the POPs, COFs and SiCOFs, these amorphous reticular materials were coined PolySilicate Porous Organic Polymers (PSiPOPs). The here presented PSiPOPs are isoreticular variants, altering only the length of the carbon chain of the diol from ethylene glycol (C2) to 1,5-pentanediol (C5), and 1,7-heptanediol (C7). Due to their structural similarity to SiCOF and (Si)POP materials but exploiting their flexible nature, PSiPOP materials are expected to have applications in the field of high-pressure gas storage, catalysis, and separation processes.

**Materials and Methods**

All spherosilicate (dimethyl silyl chloride & dimethylsilyl pyridinium chloride) and PSiPOP syntheses were performed in an inert atmosphere using a standard schlenkline set-up in combination with a coldtrap and external heating bath.

**Chemical and Reagents.** Tetrabutylammonium hydroxide (Acros organics, 40 wt% in $H_2O$), ammonium hydroxide (CHEM-LAB, 29 wt% in $H_2O$), TEOS (Acros organics, 98%), tetrahydrofuran (Acros organics, 99.8%, extra dry over molecular sieves, stabilized, acroseal), dimethyldichlorosilane (Acros organics, >99%, acroseal), triethylamine (Sigma-Aldrich, >99%), N,N-dimethylformamide (Sigma-Aldrich, 99.8%, anhydrous), pyridine (Sigma-Aldrich, 99.8%, anhydrous), ammonia hexafluorophosphate (Acros organics, 99,5%) ethylene glycol (Sigma-Aldrich, 99.8%, anhydrous), 1,5-pentanediol (Sigma-Aldrich, >97%), 1,7-heptanediol (Sigma-Aldrich, 97%), ethanol (Fischer scientific, 99.8%), NaOH (Fischer scientific, analytic reagent grade, pearls), and HCl (Acros organics, 37% in $H_2O$) were used as purchased, without any purification. Water refers to the use of milli-q water.

**Tetrabutylammonium cyclosilicate hydrate (TBA-CySH) synthesis.** Tetrabutylammonium cyclosilicate hydrate synthesis is based on the work of Smet et al. (2017).[55] TEOS (139 ml) was added dropwise to a stirred solution of tetrabutylammonium - (140.52 ml) and ammonium hydroxide (222.47 ml), to prevent gelation. After 48h, a white suspension was filtered (whatman qualitative filter paper grade 5, 90mm diameter) and washed with water. After drying at ambient conditions, TBA-CySH crystals were recovered.

**Octakis (dimethylsilyl chloride) spherosilicate synthesis.** TBA-CySH was dried under vacuum, using a standard schlenkline set-up, at room temperature for 72h in a 100 ml, 2-neck round-bottom flask, sealed with rubber stoppers and high-vacuum silicone grease. THF and dimethyldichlorosilane was added to dried TBA-CySH crystals according to a molar ratio of 130/36/1. After 1h of reaction, THF and excess of silane were evaporated at vacuum at room temperature for 2h, yielding a white dimethylsilyl chloride spherosilicate powder. These spherosilicates were dissolved in the same amount of THF yielding a spherosilicate solution (1) suitable for further synthesis.

**Dimethylsilyl pyridinium chloride spherosilicate (Si-N transition state) synthesis.** Ammonia hexafluorophosphate and pyridine were added in that order with a molar ratio of 8/8/1 to (1). After 16h of reaction, THF is evaporated at room temperature and a solid phase was collected in an inert atmosphere (minimizing the contact with any source of water) and characterized.

**PolySilicate Porous Organic Polymer (PSiPOP) synthesis.** It should be noted that octakis (dimethylsilyl chloride) are sensitive to moisture and impurities, and can influence the final composition of the PSiPOP materials (See discussion SI-2.1. for more details). For each diol linker, the synthesis proceeded as follows. Triethylamine, a catalytic amine (pyridine or DMF), and a diol linker were added chronological to (1). After 6h of reaction, the excess of reagents was evaporated at vacuum and 100°C for 16h, yielding PSiPOP powders. The volume of the used reagents are specified per diol linker in table 1. The excess of each reagent is calculated on a molar base to the number of Si-Cl bonds present in the spherosilicate solution (1).

**Table 1.** Reagents and their volume added to the spherosilicate solution for each of the 3 diol linkers.

| Diol linker | Volume (ml) | Excess (%) |
|---|---|---|
| ethylene glycol | | |
| Triethylamine | 1.5 | 20 |
| DMF | 0.35 | -50 (deficit) |
| ethylene glycol | 0.5 | 100 |
| | | |
| 1,5-pentanediol | | |
| Triethylamine | 1.5 | 20 |
| Pyridine | 0.73 | 0 |
| 1,5-pentanediol | 1 | 100 |
| | | |
| 1,7-heptanediol | | |
| Triethylamine | 1.5 | 20 |
| DMF | 0.7 | 0 |
| 1,7-heptanediol | 1.25 | 100 |

**Detemplation of PSiPOP materials.** PSiPOP materials with 1,5-pentanediol and 1,7-heptanediol linkers were calcined in a u-tube flow oven under $N_2$ atmosphere at 300°C with a rate of 1°C/min. The calcination temperature was kept for 2h before gradually being cooled to room temperature. Ethylene glycol PSiPOP was detemplated with 2 wash steps, using consecutively water and ethanol. After each step, PSiPOP powder was recovered with centrifugation (15 min at 15 000 rpm) or filtration and dried at ambient conditions after the last step.

**Hydrolytic stability.** Stability of the PSiPOP materials was analyzed by immersing the powder in an aqueous HCl and NaOH solution of pH 1 & 3 and pH 12 respectively. Vigorously stirring with a magnetic stirrer ensured sufficient contact between the solution and PSiPOP particles. After stirring for one week at room temperature, PSiPOP powder was recovered again by filtration (whatman qualitative filter paper grade 5, 90mm diameter) and washed with excessive water. The dry powder was characterized with solid-state $^{29}$Si MAS NMR.

**Characterization of PSiPOP materials.** Powder XRD patterns were recorded in transmission mode on a STOE stadi P diffractometer (Darmstadt, Germany) equipped with an image plate detector. Thermogravimetric analysis under $N_2$ atmosphere were done using a TGA Q500 apparatus (Mettler Toledo). At a rate of 10°C/min, each sample was heated from room temperature to 900°C before gradually cooled down to ambient conditions. Scanning electron microscopy images were taken with a Nova NanoSEM 450 (FEI Eindhoven) with a voltage of 2kV. Pulverized powder materials were dispersed on carbon tape without any coating. BET surfaces and pore size distributions were calculated using the BET and BJH method, respectively, according to nitrogen adsorption and desorption isotherms recorded at -196°C using a Micrometrics TriStar 3000 apparatus. Samples were heated up to 200°C at a rate of 5°C/min for 1h before degassed and cooled to 100°C. This temperature was kept constant until the measurements started.

All solid-state NMR experiments were done in Bruker 500 Avance III (11.4 T) and Bruker 800 Neo (18.8 T) spectrometer (with respective $^{29}$Si Larmor frequencies of 159.18 MHz and 99.51 MHz), with 4mm triple resonance MAS probes. $^1$H NMR experiments were done with 8 number of scans and 1 s recycle delay at 500 MHz and 4 s delay at 800 MHz. 64 transients were collected for cross-polarization (CP) MAS experiments at 15 kHz MAS with a contact time of 6.5 ms. All experiments were carried out with $^1$H decoupling of strength 55 kHz achieved using SWf-SPINAL sequence.[58] The $^{29}$Si chemical shifts were referenced against the resonances of $Q_8M_8$, a secondary reference of tetramethylsilane (TMS). $^1$H-$^1$H double quantum-single quantum (DQ-SQ) (800

MHz) experiments were done with BABA pulses of strength 50 kHz.[59] 1024 slices were collected with an increment of 66.67 µs (rotor synchronized). The 2D correlation experiments were done $^1$H-$^{29}$Si CP under MAS. $^{29}$Si CP-BABA DQ-SQ (500 MHz) experiment was done at 7 kHz MAS using a CP contact of 6.5 ms and $^{29}$Si pulses of strength 70 kHz. 600 slices were collected with an increment of 33.33 µs, with 736 transients recorded for each slice. $^{29}$Si CP-RFDR (500 MHz) experiment was done at 20 kHz MAS using a CP contact of 6.5 ms and $^{29}$Si pulses of strength 70 kHz. 700 slices were collected with an increment of 66.67 µs (rotor synchronized), with 416 transients recorded for each slice at an RFDR mixing time of 90 ms. The 2D $^1$H-$^{29}$Si CP-HETCOR (800 MHz) experiment was carried out using a CP contact time of 2 ms. A recycle delay of 4 s was used for 160 scans per slice of the 2D experiment. With an increment of 25 µs, 640 slices were recorded.

**Results and Discussion**

The synthesis of dimethylsilyl chloride spherosilicates has previously been reported by Smet et al. (2017).[55] As shown in Figure 3a, its successful implementation can be verified using solid-state $^{29}$Si *magic angle spinning* (MAS) nuclear magnetic resonance (NMR) spectroscopy. Octakis (dimethylsilyl chloride) spherosilicates exhibits $^{29}$Si resonances (Figure 3a) in 2 regions, resulting from the presence of dimethylsilyl chloride groups ($D^1$ Si; O-**Si**(CH$_3$)$_2$Cl; ~ 9 ppm) and Si atoms of the silicate D4R cube ($Q^4$ Si; ~ -110 ppm).[55] Integrating the resonances in these regions yields a $D^1$ (Si-Cl)/$Q^4$ ratio close to 1, indicating all cubes have been functionalized without any silicone linkage between 2 corners of adjacent cubes. The splitting of the $D^1$ and $Q^4$ resonances has not yet been explained. The spectrum reveals an impurity in the $D^2$ region (O-**Si**(CH$_3$)$_2$O ; ~ -18 ppm). The absence of any correlation between the $D^2$ and the $D^1$ or $Q^4$ resonances in the $^{29}$Si-$^{29}$Si radiofrequency driven dipolar recoupling (RFDR) spectrum (Figure 3b), while the $D^1$ and $Q^4$ resonances correlate as expected, confirms these $D^2$ atoms reside in a separate impurity. This impurity is identified as a dimethylsilicone polymer originating from self-polymerisation of chlorosilane hydrolyzed with traces of water.

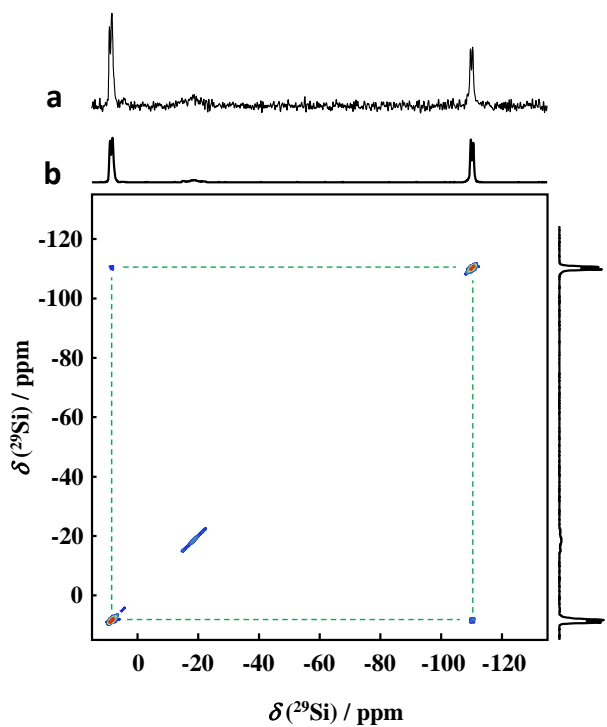

**Figure 3**. a. 1D $^1$H decoupled $^{29}$Si MAS NMR spectrum and b. 2D $^{29}$Si-$^{29}$Si radiofrequency driven dipolar recoupling (RFDR) NMR spectrum of octakis (dimethylsilyl chloride) spherosilicates.

As first documented in 1957 by Allen et al.,[60] chlorosilanes readily react with alcohols. Referred to as 'Corey's method', this reaction has been applied numerous times as a protection strategy for hydroxyls in multifunctional molecules.[61] While Corey and co-workers used *tert*-butyldimethylsilyl chloride as effective protection agent for secondary alcohols, by now many variations of alcohol substrates (primary, secondary, and tertiary), silyl protecting groups (chlorosilanes, hydrosilanes, and alkylsilyl triflates), catalysts (Lewis base amines and n-oxides), and solvents have been reported.[62–71] In general, a Lewis base amine catalyst activates the chlorosilane by forming a silylated quaternary ammonium transition state. Subsequent nucleophilic attack of the hydroxyl group yields a silylether. Auxiliary amines such as triethylamine (Et$_3$N) close the catalytic cycle by regenerating the catalyst through proton transfer.[72] Whereas this method in its original application is exclusively used to protect alcohol functions in organic synthesis, it offers opportunities for further functionalizing octakis (dimethylsilyl chloride) spherosilicates and in extension for reticular material synthesis. A potential reaction mechanism for such synthesis is shown in Figure 4, depicting a pyridine-catalyzed reaction with a generic diol (OH-(CH$_2$)$_n$-OH).

Evidence for the intermediate formation of a silylated quaternary ammonium transition state (Si-N) with pyridine was obtained by solid-state $^{29}$Si MAS NMR (Figure 4). For this experiment, pyridine and ammonium hexafluorophosphate were added to dimethylsilyl chloride spherosilicate dissolved in tetrahydrofuran (THF). Ammonium hexafluorophosphate was added as a weakly coordinating anion to remove the chloride via an ammonium chloride salt. After 16h of reaction, excessive reagents were evaporated at room temperature, resulting in a white powder. Compared to the octakis (dimethylsilyl chloride) spherosilicates spectrum (Figure 3a), the 1D $^{29}$Si MAS NMR spectrum of the octakis (dimethylsilyl pyridinium chloride) exhibited two additional resonances at -3.6 ppm & -6.4 ppm, associated to the formation of a Si-N linkage (Figure 4). The chemical shift occurs at higher chemical shift as compared to D$^2$-coordinated Si atoms because of the lower electronegativity of nitrogen (3.04) as compared to oxygen (3.44). As for the splitting in the D$^1$ and Q$^4$ resonances, the splitting of resonance associated with the Si-N intermediate has not yet been explained. The presence of pyridine was further confirmed using $^1$H MAS NMR, revealing pyridine protons between 6 and 9 ppm (Figure SI-1).

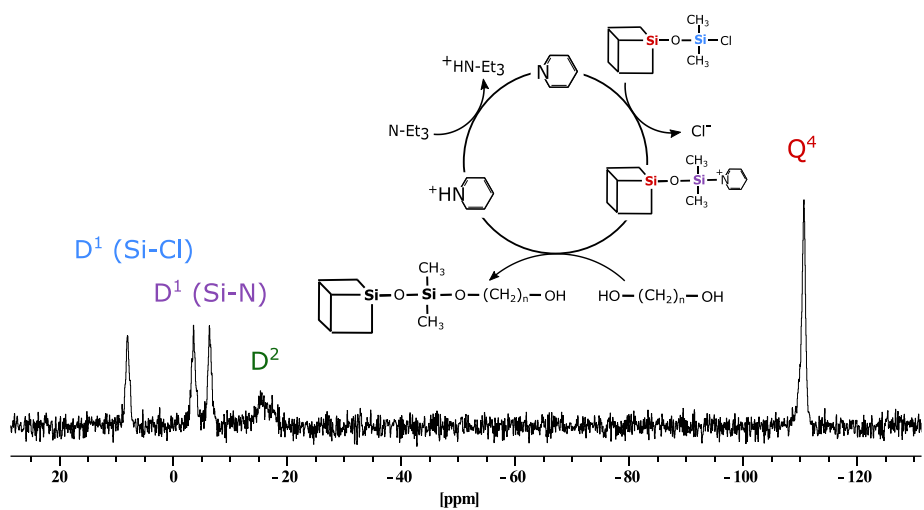

**Figure 4.** Reaction mechanism where pyridine catalyzes the reaction of octakis (dimethylsilyl chloride) spherosilicates with linear, aliphatic α,ω-alkanediols. The reaction proceeds via a Si-N intermediate. Triethylamine (Et$_3$N) regenerates the pyridine. Proof of the Si-N intermediate is given by $^1$H decoupled $^{29}$Si solid-state MAS NMR spectrum, where the 2 resonances at -3.4 & - 6.4 ppm are ascribed to the Si species of the intermediate.

Performing a similar reaction, adding triethylamine and 1,5 pentanediol to a sealed reaction flask containing dimethylsilyl chloride spherosilicates dissolved in THF containing the catalytic pyridine, yielded the first PSiPOP material (Table 1). After 6h of reaction at room temperature and evaporation of excess of THF and high-boiling reagents at 100°C under vacuum (< 1 mbar), a white, porous 3D reticular material was recovered. Since x-ray diffraction demonstrated the exclusive presence of amorphous material (Figure SI-2 – SI-4), the reticular network was elucidated using an NMR crystallography approach (Figure 5).[73–76]

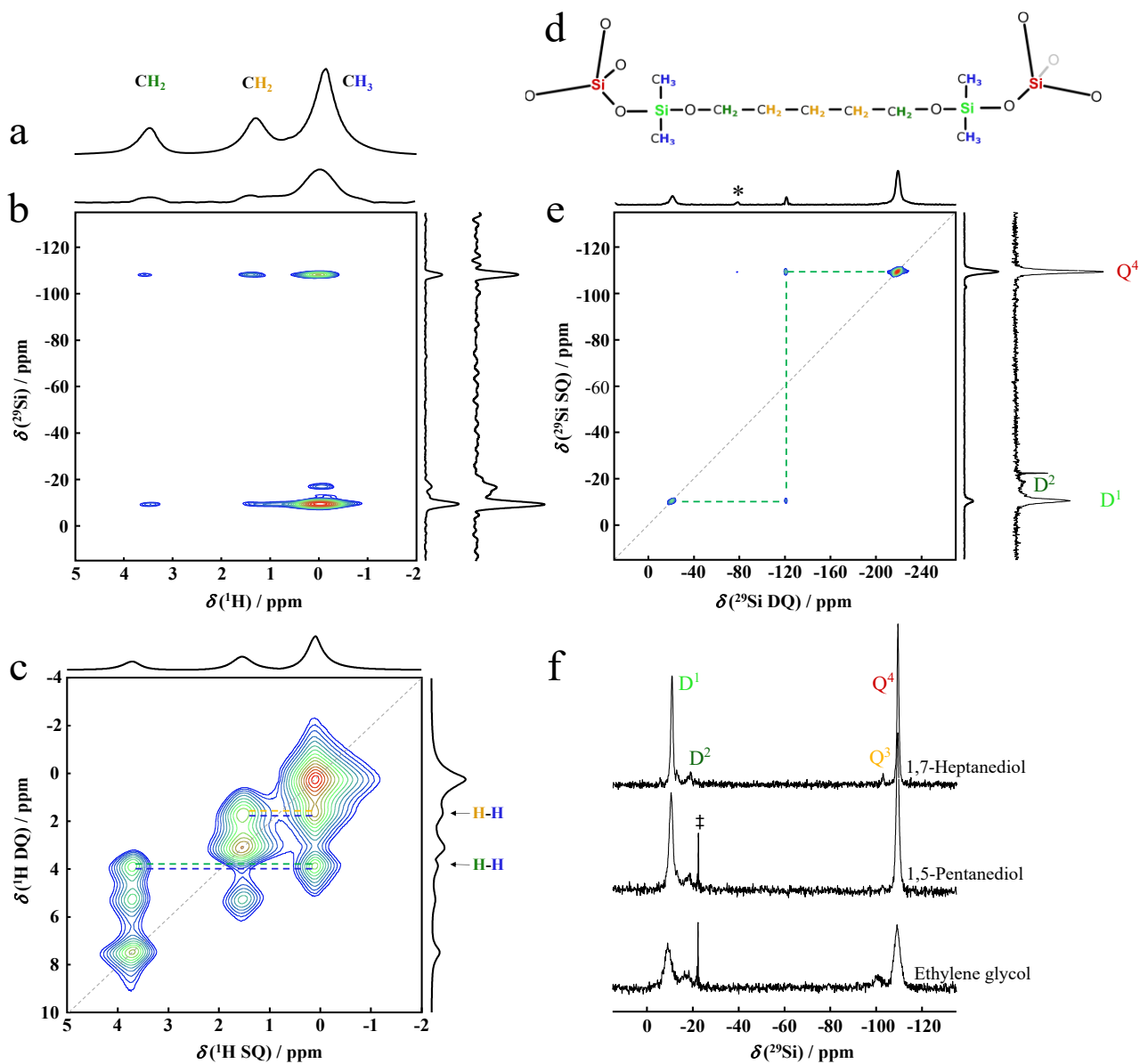

**Figure 5**. Solid-state NMR crystallography analysis of a PSiPOP material with 1,5-pentanediol linkers. **a**. 1D direct excitation $^1$H MAS NMR spectrum; **b**. 2D $^1$H -$^{29}$Si HETCOR spectrum; **c**. 2D $^1$H -$^1$H dipolar DQ-SQ correlation spectrum. Cross-correlation between the $^1$H of the dimethyl silyl groups and the methylene protons are marked with dashed lines in the colours used to mark the respective protons in the structure model in Figure 5d; **d.** chemical structure model of the pentanediol PSiPOP; **e.** 2D $^{29}$Si - $^{29}$Si DQ-SQ spectrum with $^1$H-$^{29}$Si CPMAS excitation. Autocorrelations are marked on the diagonal line (grey dash); cross-correlations appear off-diagonal and are indicated with the green dashed line. The * marks a spinning side-band; **f.** 1D $^1$H decoupled $^{29}$Si single-pulse MAS NMR spectra of PSiPOP materials linked with heptanediol (top), pentanediol (middle), and ethylene glycol (bottom) respectively. The sharp resonance at 22.4 ppm, indicated with ‡, is assigned to a small contamination with silicone grease, used to seal the glass reaction vessel airtight.

As revealed by the 1D $^{29}$Si MAS NMR spectrum (Figure 5f), the reticular material contains similar $D^1$ and $Q^4$ resonances as observed for the dimethylsilyl chloride spherosilicate nodes (Figure 3a). Incorporation of the diol linkers induced a change in chemical shift of the $D^1$ Si atoms, moving this resonance from +9 to -9 ppm. As expected from the structure model (Figure 5d), the $^1$H-$^{29}$Si HETCOR spectrum (Figure 5b) correlates the two types of -CH$_2$- protons of the pentanediol linker with both $D^1$ and $Q^4$ Si atoms, while the $^{29}$Si-$^{29}$Si DQ-SQ measurement with $^1$H-$^{29}$Si CPMAS excitation (Figure 5e) confirms the correlation between the $D^1$ and $Q^4$ Si atoms. This combination of 2D NMR spectra readily proves the incorporation of diol linkers in the reticular network. Further evidence of the incorporation was provided by $^1$H-$^1$H DQ-SQ MAS spectroscopy (Figure 5c). Consistent with the assignment of the 1D $^1$H MAS NMR spectrum (Figure 5a), the $^1$H-$^1$H DQ-SQ spectrum readily reveals self-correlations (on-diagonal) for the methyl substituents of the $D^1$ Si and for the two types of -CH$_2$- protons in the pentanediol linker (Figure 5c & Discussion SI-2.2.). The off-diagonal DQ correlations between the -CH$_3$ and the two types of -CH$_2$- protons again reveals through-space correlation of these nearby positioned protons, a situation resulting from the successful generation of a Si-O-C linkage between the spherosilicate and the diol linker (Discussion SI-2.3.). The relative concentrations of the $^{29}$Si resonances in the 1D $^{29}$Si MAS NMR spectra, as derived from the decomposed spectra can be found in the Supporting Information as Table SI-1 and Figure SI-5, respectively. This efficient PSiPOP synthesis using an extended Si$_{16}$ spherosilicate node linked up with aliphatic α,ω-alkanediols via a Si-O-C linkage broadens the variety of silicate nodes and synthesis pathways for reticular SiCOF and SiPOP materials. Previous SiCOF reports implementing D4R-based nodes, remained limited to a Si$_8$ octakis (phenyl) D4R POSS as node, connected to the linker via a Si-C bond and with a very low yield of the final hybrid material.[44]

Following the successful synthesis of this pentanediol-linked PSiPOP, an isoreticular PSiPOP series was generated, implementing ethylene glycol and 1,7-heptanediol as alternative linkers. Instead of pyridine,[64,70] for these syntheses DMF,[62,77] was used to promote the reaction of octakis (dimethylsilyl chloride) spherosilicates with the alcohols. While this demonstrates that any catalyst with a proven track record in 'Corey's method', can probably be used for these syntheses, evaluating their relative performance falls outside the scope of the present work.

Figure 5f compares the 1D $^1$H decoupled $^{29}$Si direct excitation MAS NMR spectra for the isoreticular series. Independent of the length of the carbon chain of the diol linkers, similar features are observed in the 1D $^{29}$Si MAS NMR spectra: (i) Efficient occupation of all corners with dimethyl silyl groups, represented by the $Q^4$ signal at -109.3 ppm. (ii) $D^1$ coordinated Si atoms (-10 ppm) resulting from the incorporation of diol molecules in the reticular materials. The minor concentration of $Q^3$ Si atoms (~ -100 to -102 ppm) results from unoccupied corners of the silicate cube. The slight increase of unoccupied corners with a shorter linker is probably a result of steric hindrance during formation of the Si-N transition state and the subsequent diol addition. The $D^2$-coordinated silicon impurity (-18 ppm) results from a minor concentration of dimethyl silicone oligomers formed by the self-polymerisation of chlorosilane with traces of water (Discussion SI-2.1.).

The thermal stability of the hybrid materials was probed with thermogravimetric analysis in inert ($N_2$) atmosphere (Figure SI-6). Decomposition of the amines salts (i.e. tetrabutylammonium chloride (TBA-Cl) and triethylammonium chloride ($Et_3$N-HCl)) induces a first weight loss that occurs around 190°C. A second weight loss centered around 490°C results from the decomposition of the organic moieties in the linker between 2 silicate cubes: (-O-Si(**CH₃**)₂)-O-(**CH₂**)ₙ-O-Si(**CH₃**)₂-O-), with n = 5, or 7 for respectively pentanediol, and heptanediol. For ethylene glycol (n=2), organic moieties decompose at lower temperatures (around 250°C). The Si-O-C bond in diol-linked orthosilicate gels has been shown to easily withstand temperatures exceeding 300°C in an inert atmosphere (Ar).[78] Dimethylsilyl groups have been demonstrated to exhibit a high thermal stability in the related PSS-2 materials.[56] The reported PSiPOP materials are therefore stable up to 400°C in inert atmosphere. Thermal analysis as well as $^1$H direct excitation MAS (Figures SI-8, 10, & 12) and $^1$H-$^{13}$C CPMAS spectroscopy (Figures SI-9, 11, & 13 (left)) indicate the presence of amine salts in the as-synthesized PSiPOP materials. For the materials with pentanediol and heptanediol linkers, removal of these salts by calcination in $N_2$ at 300°C is straightforward and the thermal stability is maintained after the calcination (Figure SI-7). The removal of the mobile salt fraction is readily demonstrated by the disappearance of the sharp resonances from $^1$H NMR and $^{13}$C spectra of the as-synthesized material (Figures SI-8 – SI-11). Stability of the framework is also confirmed as the broader resonances remain identical before and after calcination (Figures SI-8 – SI-11). For the ethylene glycol-linked material, removal of the salts is not straightforward as no thermal equilibrium between the two decomposition steps was observed in the TGA measurement (Figure SI-6). Consequently, a procedure with consecutive water and ethanol rinsing steps was used. This method also proved to be effective for removal of the amine salts without damaging the framework structure (Figures SI-12 and 13). Figure SI-14 shows SEM images of all PSiPOP materials before and after removal of the amine salts.

As previously observed for the POSiSil materials, built up from D4R cubes with silicone linkers, also in these PSiPOP materials, the micropore volume appears to be inaccessible in liquid nitrogen conditions, explaining the rather low pore volume probed by $N_2$ physisorption (Figure SI-15). Likely, this is due to a deformation or collapse of the flexible structure at these low temperatures.[56]

Asides thermal stability, also the hydrolytic stability of the materials was investigated in aqueous conditions at pH 1, 3, and 12 by soaking the material for 7 days in the respective solutions at room temperature, followed by filtration, rinsing with water, and drying at 60°C. The resulting dried material was characterized using direct excitation $^1$H MAS, $^1$H-$^{13}$C CP MAS, and $^1$H-$^{29}$Si CPMAS NMR. Comparison of the spectra presented in Figures SI-16 – SI-18 to those of the original material readily demonstrates the hydrolytic stability of the siloxy bonds in the materials. This feature could be related to the synergy of steric hindrance and hydrophobicity of octakis (dimethylsilyl) silicate D4Rs and the diol linkers. Water is repelled from the surface and the pore space, and the siloxy bonds are maintained throughout the reticular network. This is in analogy with the increased hydrolytic stability of organoalkoxysilanes with more bulky organic functionalities.[79]

**Conclusion**

In conclusion, the research presented in this study demonstrates the successful synthesis and characterization of a novel class of reticular materials termed PolySilicate Porous Organic Polymers (PSiPOPs). These PSiPOP frameworks feature $[Si_8O_{20}][Si(CH_3)_2]_8$ nodes interconnected by α,ω-alkanediol linkers, with specific examples being ethylene glycol, 1,5-pentanediol, and 1,7-heptanediol. The synthesis proceeds through a two-step process, involving the conversion of sacrificial hydrogen-bonded tetrabutylammonium $Si_8O_{20}$ cyclosilicate crystals into silyl chloride terminated spherosilicate nodes, followed by the interconnection of these nodes using aliphatic α,ω-alkanediol linker molecules. The macroscopic properties of the synthesized PSiPOP materials showcase thermal stability up to 400°C and chemical stability across a pH range from 1 to 12. Additionally, $N_2$ physisorption analysis has revealed a secondary mesopore structure, hinting at the potential to engineer hierarchical materials using soft templates in the future. The molecular-level structure elucidation of these PSiPOP materials has been achieved through an NMR crystallography approach, involving a combination of 1D and 2D $^1$H and $^{29}$Si solid-state MAS NMR spectroscopy experiments.

In essence, this research introduces a promising avenue in the realm of reticular porous materials, expanding the field beyond the well-established metal organic frameworks (MOFs), covalent organic frameworks (COFs) and Porous Organic Polymers (POPs). The here presented synthesis of PSiPOPs hints at the possibility for implementing more rigid linker molecules and extending their functionalisation into a more polar chemistry. This could yield crystalline materials and provide options for tailoring their properties for various applications such as high-pressure gas storage, catalysis, and separation processes. The detailed insights gained into the molecular structure and mechanism of formation of the new PSiPOP family of materials paves the way for further advancements in their design and optimization.

ACKNOWLEDGMENT

E.B. acknowledges joint funding by the Flemish Science Foundation (FWO; G083318N) and the Austrian Science Fund (FWF) (funder ID 10.13039/501100002428, project ZeoDirect I 3680-N34). J.A.M acknowledges the Flemish government for long term structural funding (Methusalem) and the European Research Council (ERC) for an Advanced Research Grant under the European Union's Horizon 2020 research and innovation program under grant agreement No. 834134 (WATUSO). NMRCoRe acknowledges the Flemish government, department EWI for infrastructure investment via the Hermes Fund (AH.2016.134) and for financial support as International Research Infrastructure (I001321N: Nuclear Magnetic Resonance Spectroscopy

Platform for Molecular Water Research). The authors acknowledge Gina Vanbutsele, Karel Duerinckx and Sreeprasanth Pulinthanathu Sree for assistance with material characterization.

## ASSOCIATED CONTENT

**Supporting Information**.

The Supporting Information is available free of charge at:….

Additional data of final PSiPOP materials (and their Si-N transition state) from XRD, NMR ($^1$H, $^{13}$C, $^{29}$Si and decompositions), TGA, SEM, and $N_2$ physisorption (PDF)

## AUTHOR INFORMATION

**Corresponding Author**

**Eric Breynaert** - Centre for Surface Chemistry and Catalysis – Characterization and Application Team (COK-KAT), KU Leuven, Celestijnenlaan 200F, Box 2461, 3001-Heverlee, Belgium; NMRCoRe - NMR/X-Ray platform for Convergence Research, KU Leuven, Celestijnenlaan 200F, Box 2461, 3001-Heverlee, Belgium; Email: Eric.Breynaert@kuleuven.be

**Authors**

**Jelle Jamoul -** Centre for Surface Chemistry and Catalysis – Characterization and Application Team (COK-KAT), KU Leuven, 3001-Heverlee, Belgium;

**Sam Smet -** Centre for Surface Chemistry and Catalysis – Characterization and Application Team (COK-KAT), KU Leuven, 3001-Heverlee, Belgium**;**

**Sambhu Radhakrishnan -** Centre for Surface Chemistry and Catalysis – Characterization and Application Team (COK-KAT), KU Leuven, 3001-Heverlee, Belgium; NMRCoRe - NMR/X-Ray platform for Convergence Research, KU Leuven, 3001-Heverlee, Belgium;

**C. Vinod Chandran -** Centre for Surface Chemistry and Catalysis – Characterization and Application Team (COK-KAT), KU Leuven, 3001-Heverlee, Belgium; NMRCoRe - NMR/X-Ray platform for Convergence Research, KU Leuven, 3001-Heverlee, Belgium;

**Johan A. Martens** - Centre for Surface Chemistry and Catalysis – Characterization and Application Team (COK-KAT), KU Leuven, 3001-Heverlee, Belgium; NMRCoRe - NMR/X-Ray platform for Convergence Research, KU Leuven, 3001-Heverlee, Belgium;

**Author Contributions**

The manuscript was written through contributions of all authors. All authors have given approval to the final version of the manuscript.

**Notes**

There are no conflicts to declare.

TABLE OF CONTENTS ENTRY

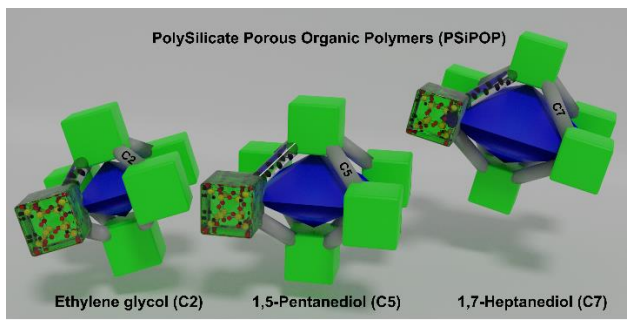

# Supporting Information

# PolySilicate Porous Organic Polymers (PSiPOPs), a new family of porous, ordered 3D reticular materials with polysilicate nodes and organic linkers

*Jelle Jamoul[a], Sam Smet[a], Sambhu Radhakrishnan[a,b], C. Vinod Chandran[a,b], Johan A. Martens[a,b], Eric Breynaert[a,b]\**

[a] Centre for Surface Chemistry and Catalysis – Characterization and Application Team (COK-KAT), KU Leuven, Celestijnenlaan 200F Box 2461, 3001-Heverlee, Belgium

[b] NMRCoRe - NMR/X-Ray platform for Convergence Research, KU Leuven, Celestijnenlaan 200F Box 2461, 3001-Heverlee, Belgium

* Email: Eric.Breynaert@kuleuven.be

**List of contents**

## SI 2. Discussions

# SI 1. Supplementary data

## SI 1.1. NMR complementing the Si-N transition

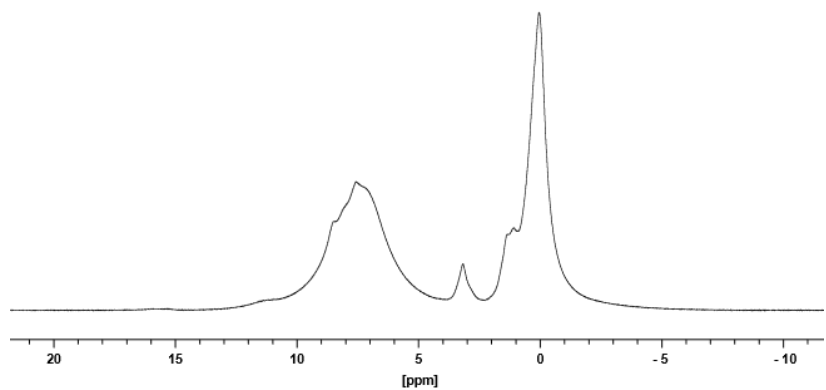

**Figure SI-1.** [1]H MAS NMR spectrum of the silylated quaternary ammonium transition state due to the reaction of dimethylsilyl chloride spherosilicates and pyridine. Pyridine's protons are in the range of 6-9 ppm. Methyl protons of the dimethylsilyl chloride functionalities are around 0 ppm, while the residual [1]H resonances are ascribed to TBA-Cl template.

## SI 1.2. XRD diffractograms complementing PSiPOP materials before and after rinsing or calcination

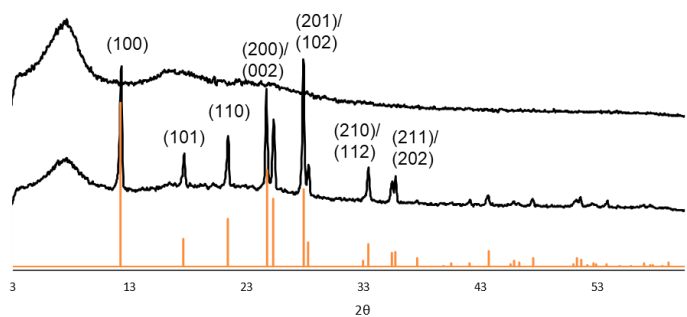

**Figure SI-2.** XRD diffractogram of ethylene glycol PSiPOP material before (middle) and after (top) a washing procedure. Reflections in the sample before a washing procedure are assigned to triethylamine hydrochloride salt. The triethylamine hydrochloride diffractogram, calculated from reference (S1), is added below and the (hkl) indices are added to most prominent reflections.

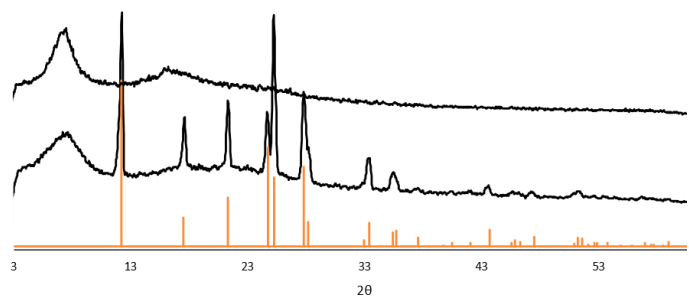

**Figure SI-3.** XRD diffractogram of 1,5-pentanediol PSiPOP material before (middle) and after (top) calcination. Reflections in the sample before calcination are assigned to triethylamine hydrochloride salt (bottom).

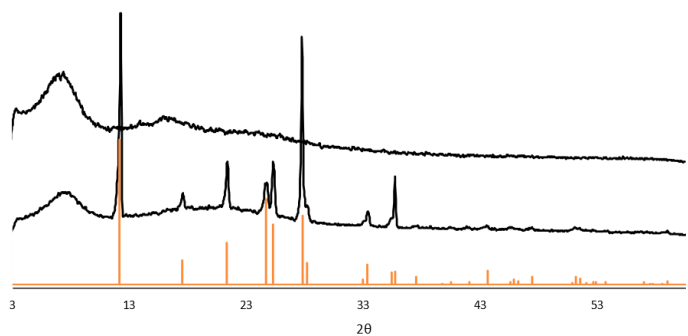

**Figure SI-4.** XRD diffractogram of 1,7-heptanediol PSiPOP material before (middle) and after (top) calcination. Reflections in the sample before calcination are assigned to triethylamine hydrochloride salt (bottom).

### SI 1.3. Decomposition of 1D $^1$H decoupled $^{29}$Si single-pulse MAS NMR spectra

**Table SI-1.** Quantitative numbers of $Q^4$, $D^1$, and $D^2$-coordinated $^{29}$Si atoms per silicate cube, according to figure SI-5.

| Si coordination | Diol linker | | |
|---|---|---|---|
|  | ethylene glycol | 1,5-pentanediol | 1,7-heptanediol |
| $Q^4$ Si | 6.5 | 7.9 | 7.7 |
| $D^1$ Si | 5 | 6.6 | 6.4 |
| $D^2$ Si | 3 | 1.9 | 1.9 |

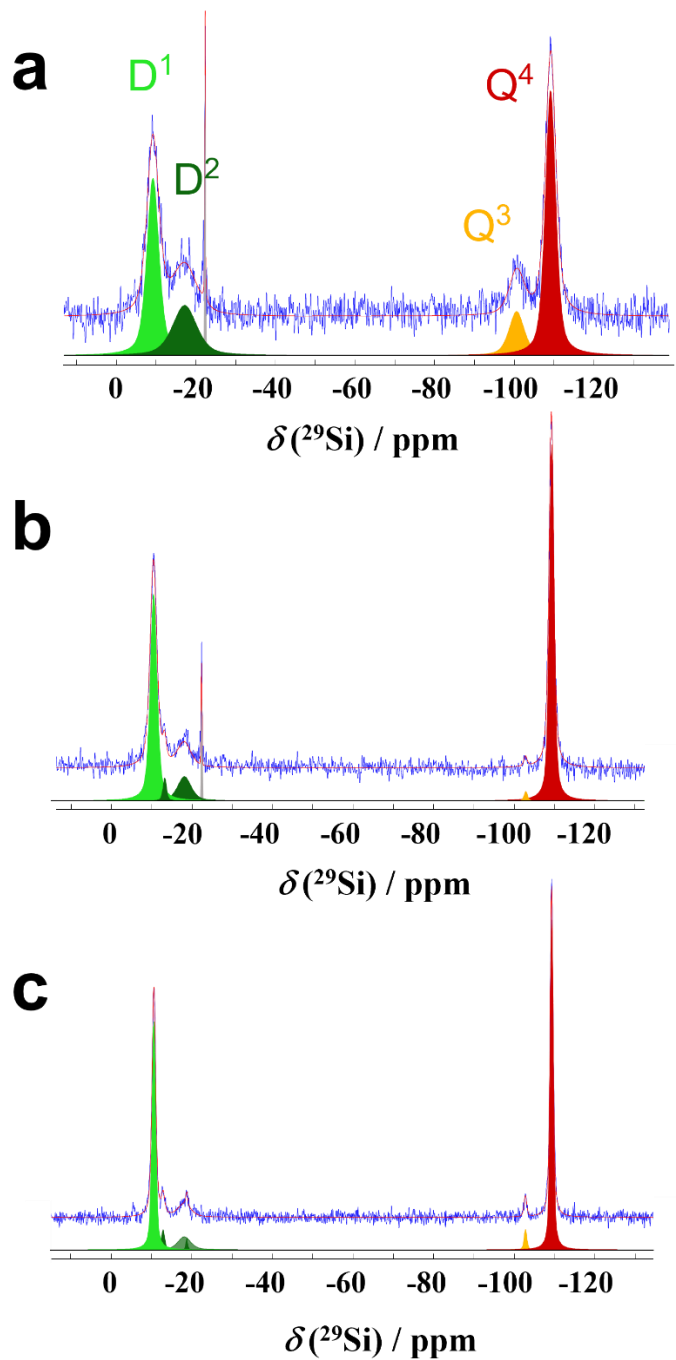

**Figure SI-5.** Decomposition of 1D $^1$H decoupled $^{29}$Si single-pulse MAS NMR spectra of PSiPOP materials with ethylene glycol (a), 1,5-pentanediol (b), and 1,7-heptanediol (c,) according to figure 5f. Specification of different Si species is assigned with different colours.

## SI 1.4. TGA analysis of PSiPOP materials in inert (N$_2$) atmosphere

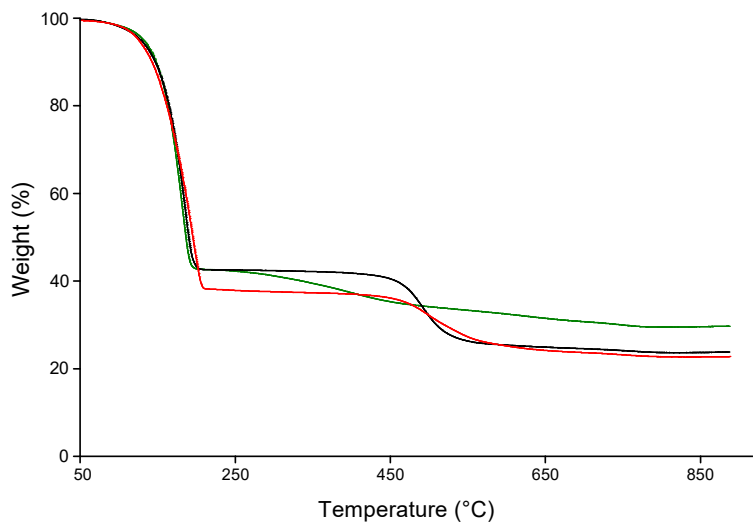

**Figure SI-6.** Thermogravimetric analysis of PSiPOP materials in inert (N$_2$) atmosphere before any treatment (calcination or wash step) with ethylene glycol linkers (green), 1,5-pentanediol linkers (black) and 1,7-heptanediol linkers (red) in N$_2$ atmosphere.

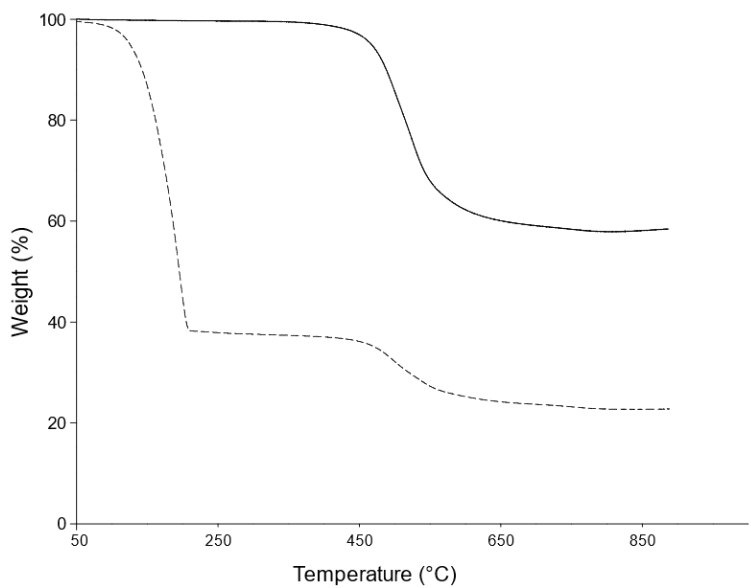

**Figure SI-7.** Thermogravimetric analysis of 1,7-heptanediol linked PSiPOP material before (dashed line) and after (full line) calcination in N$_2$ atmosphere.

**SI 1.5. NMR complementing an efficient calcination procedure**

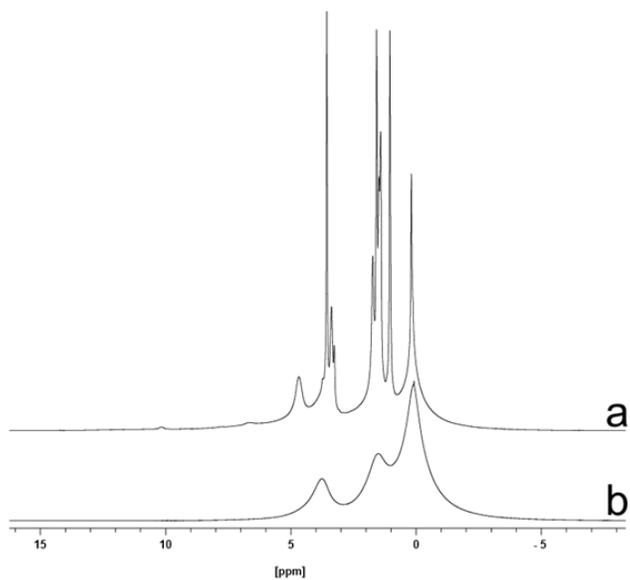

**Figure SI-8.** $^1$H spectrum of 1,5-pentanediol PSiPOP material before (a) and after (b) calcination.

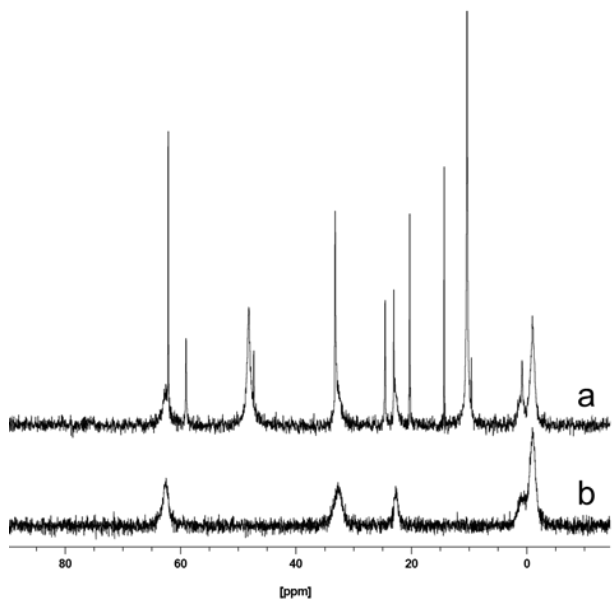

**Figure SI-9.** $^1$H → $^{13}$C CP MAS NMR spectra of 1,5-pentanediol PSiPOP material before (a) and after (b) calcination. The resonances ascribed to TBA-Cl (14.3, 20.3, 24.6 & 59 ppm) and Et$_3$N-HCl (10.3 & 48.2 ppm) disappeared upon calcination while the carbon backbone of the diol linker and the dimethylsilyl groups resonate at 22.7, 32.8, and 62.6 ppm, and at -0.9 and 0.8 ppm respectively.

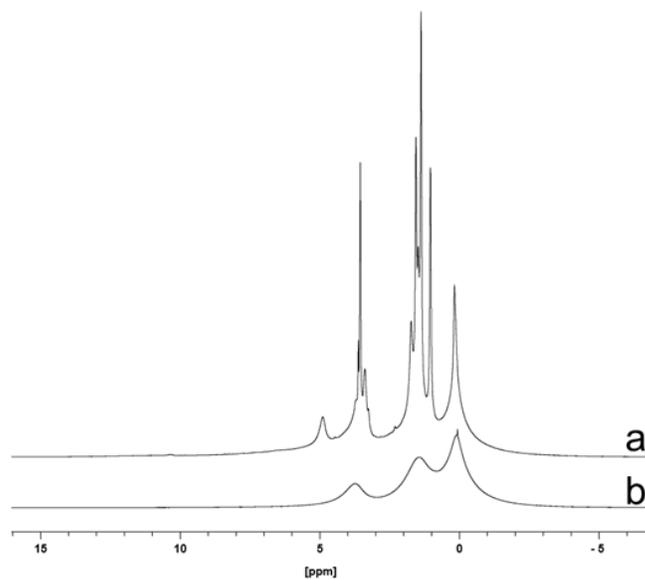

**Figure SI-10.** $^1$H spectrum of 1,7-heptanediol PSiPOP material before (a) and after (b) calcination.

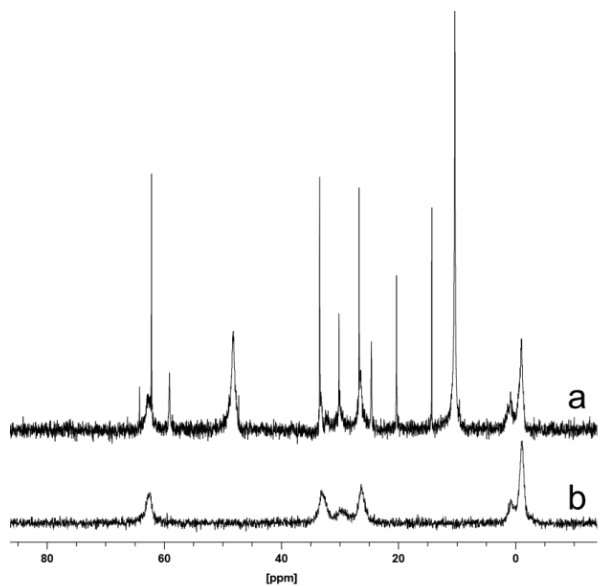

**Figure SI-11.** $^1$H → $^{13}$C CP MAS NMR spectra of 1,7-heptanediol PSiPOP material before (a) and after (b) calcination. The resonances ascribed to TBA-Cl (14.3, 20.3, 24.6 & 59 ppm) and Et$_3$N-HCl (10.3 & 48.2 ppm) disappeared upon calcination while the carbon backbone of the diol linker and the dimethylsilyl groups resonate at 26.4, 29.8, 33.1 and 62.6 ppm, and at -0.9 and 0.9 ppm respectively.

**SI 1.6. NMR complementing an efficient washing procedure**

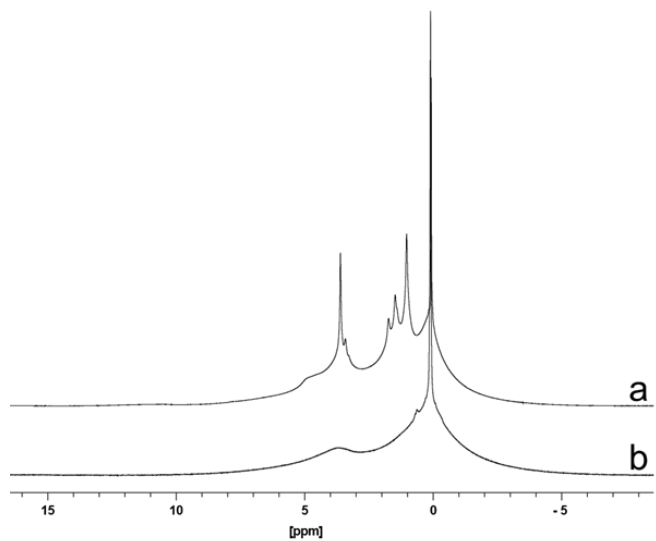

**Figure SI-12.** $^1$H spectrum of ethylene glycol PSiPOP material before (a) and after (b) a water-ethanol washing step.

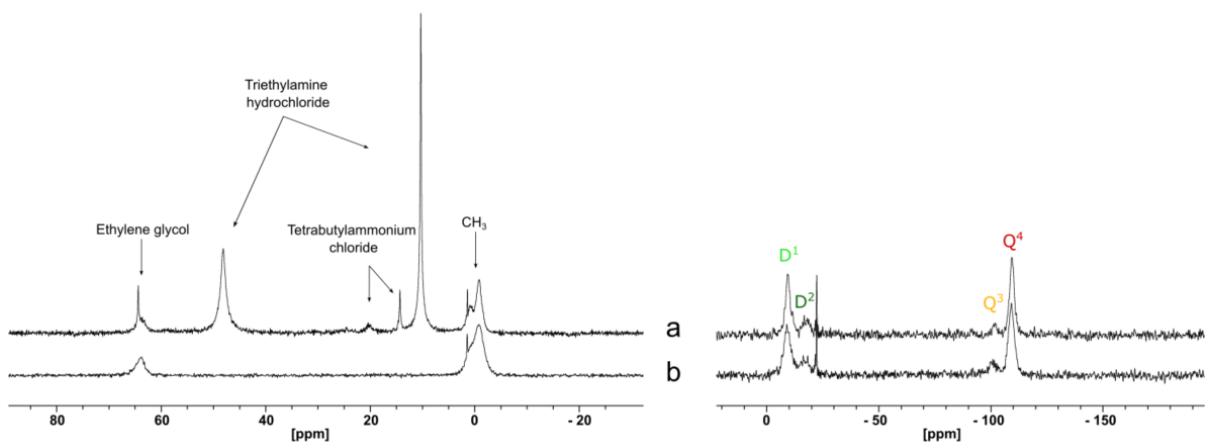

**Figure SI-13.** Left: $^1$H → $^{13}$C CP MAS NMR spectra of ethylene glycol PSiPOP material before (a) and after (b) a water-ethanol washing step. Right: Corresponding $^1$H decoupled $^{29}$Si single-pulse MAS NMR spectra.

## SI 1.7. Identifying the morphology of PSiPOP materials with SEM

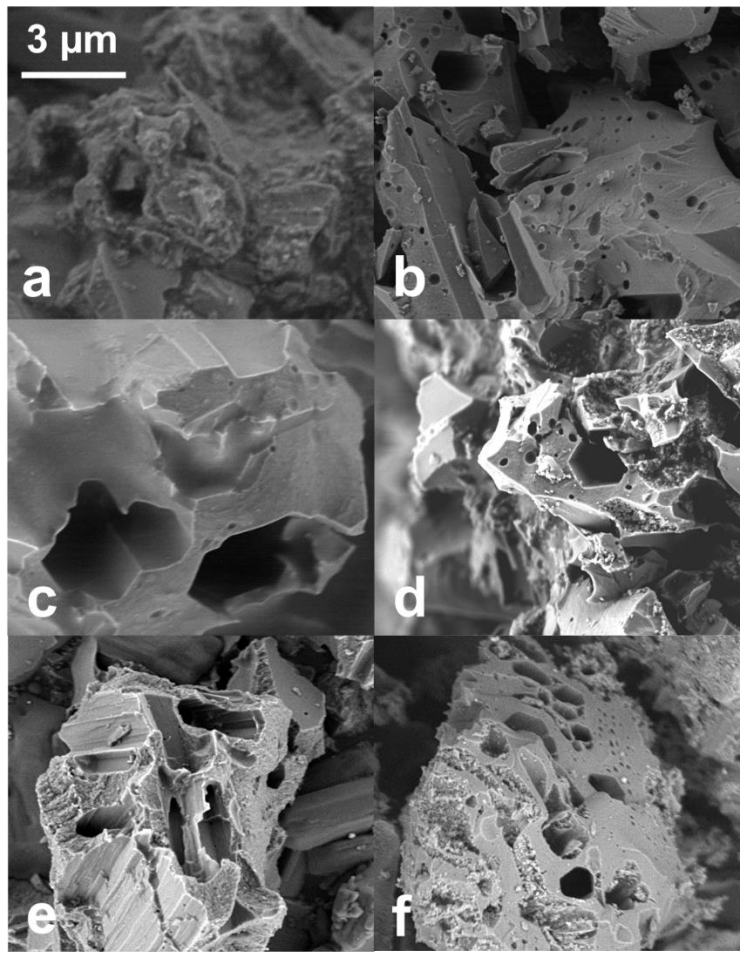

**Figure SI-14.** SEM images of PSiPOP materials before (left) and after removal of the amine salts (right) with a wash step in the case of ethylene glycol (a,b), or by means of calcination with 1,5-pentanediol (c,d) and 1,7-heptanediol (e,f) linkers.

## SI 1.8. $N_2$ physisorption data

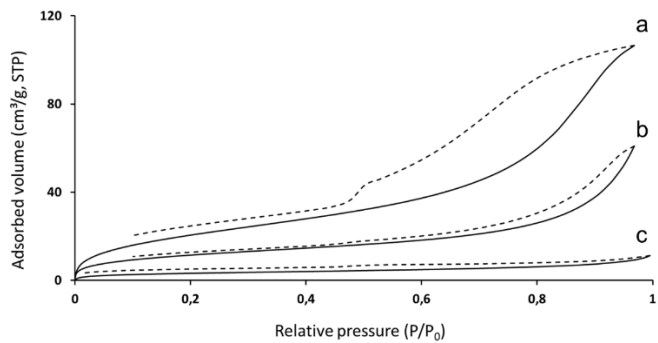

**Figure SI-15.** $N_2$ physisorption isotherms of washed or calcined PSiPOP materials with 1,7-heptanediol (a), 1,5-pentanediol (b), and ethylene glycol (c). Adsorption is given in full lines, desorption in dotted lines.

**SI 1.9. NMR investigating the stability of a PSiPOP material in alkaline and acid aqueous solutions**

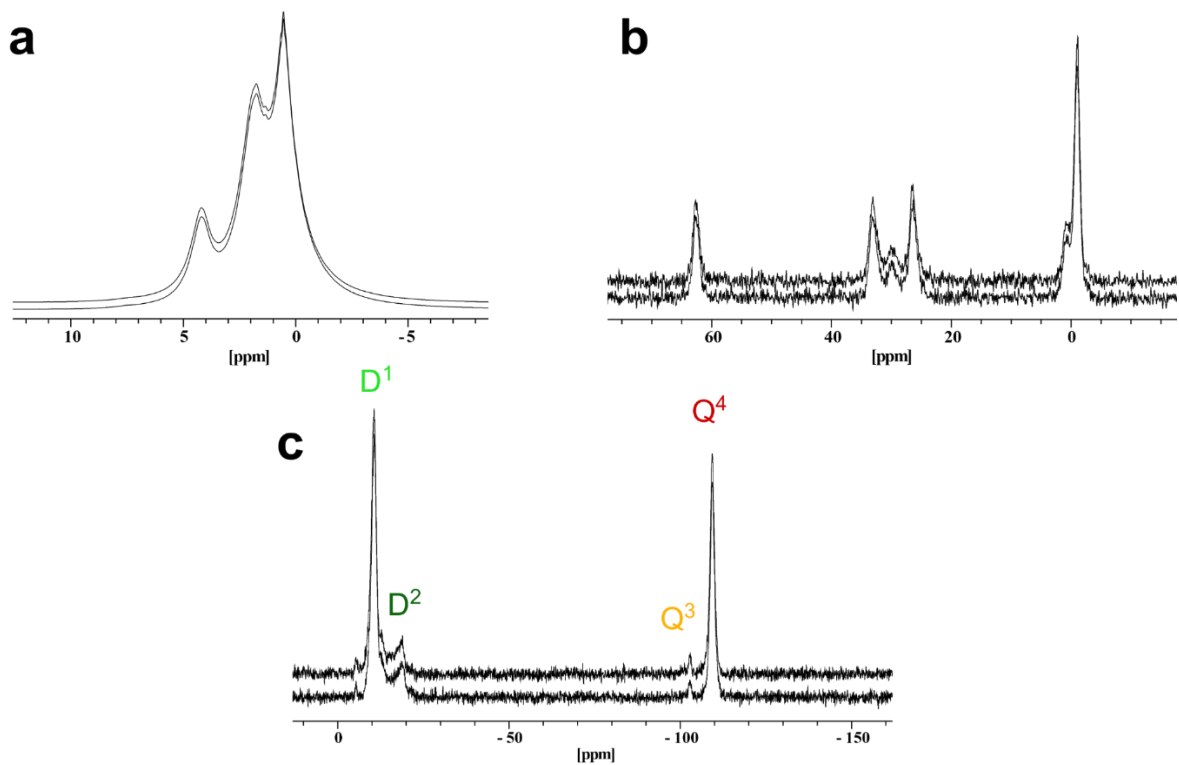

**Figure SI-16.** Investigating the stability of a PSiPOP material with 1,7-heptanediol linkers immersed in an aqueous solution of pH 12 (NaOH). $^1$H (a); $^1$H → $^{13}$C CP MAS (b), and $^1$H decoupled $^{29}$Si single-pulse NMR spectra (c) of the parent material (bottom) and after immersed in pH 12 (top). All NMR spectra are identical before and after the immersion, proving the stability of this PSiPOP material.

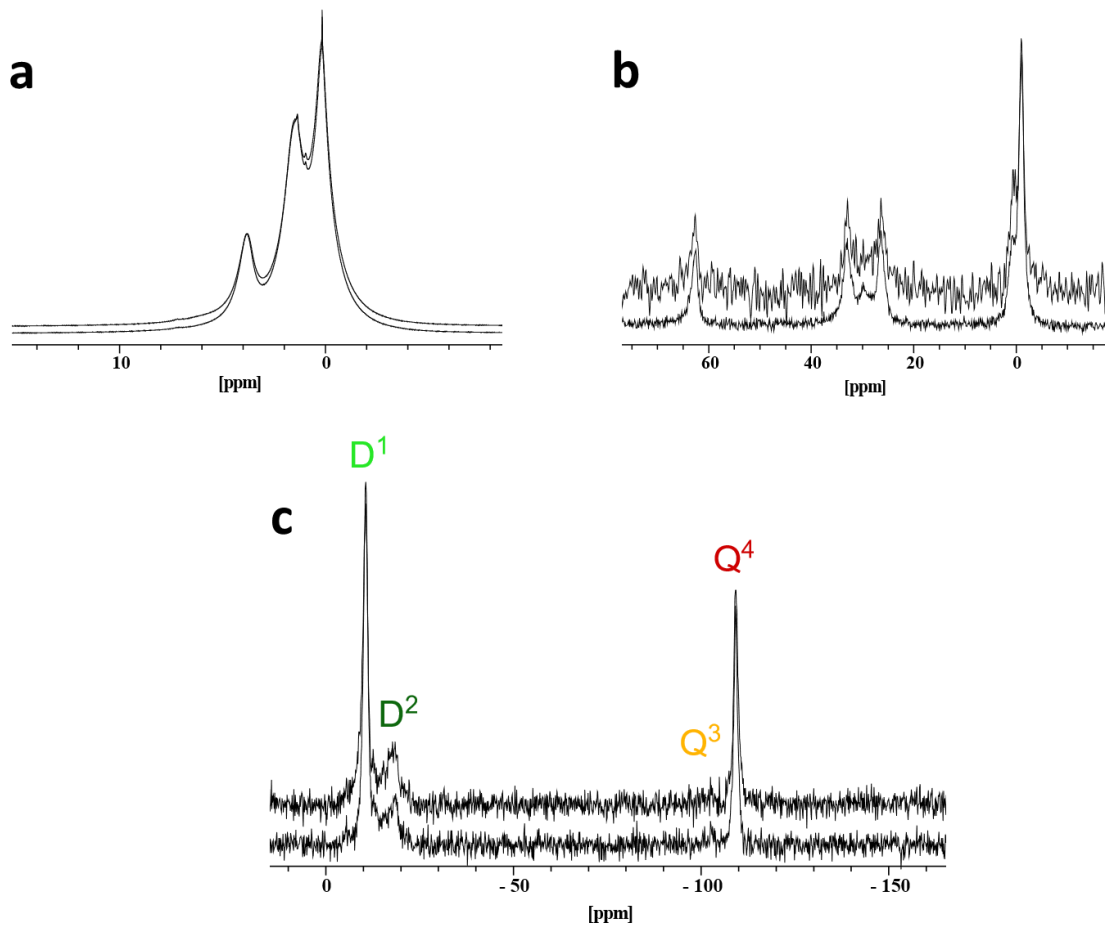

**Figure SI-17.** Investigating the stability of a PSiPOP material with 1,7-heptanediol linkers immersed in an aqueous solution of pH 1 (HCl). $^1$H (a); $^1$H → $^{13}$C CP MAS (b), and $^1$H decoupled $^{29}$Si single-pulse NMR spectra (c) of the parent material (bottom) and after immersed in pH 1 (top). All NMR spectra are almost identical before and after the immersion, proving the stability of this PSiPOP material.

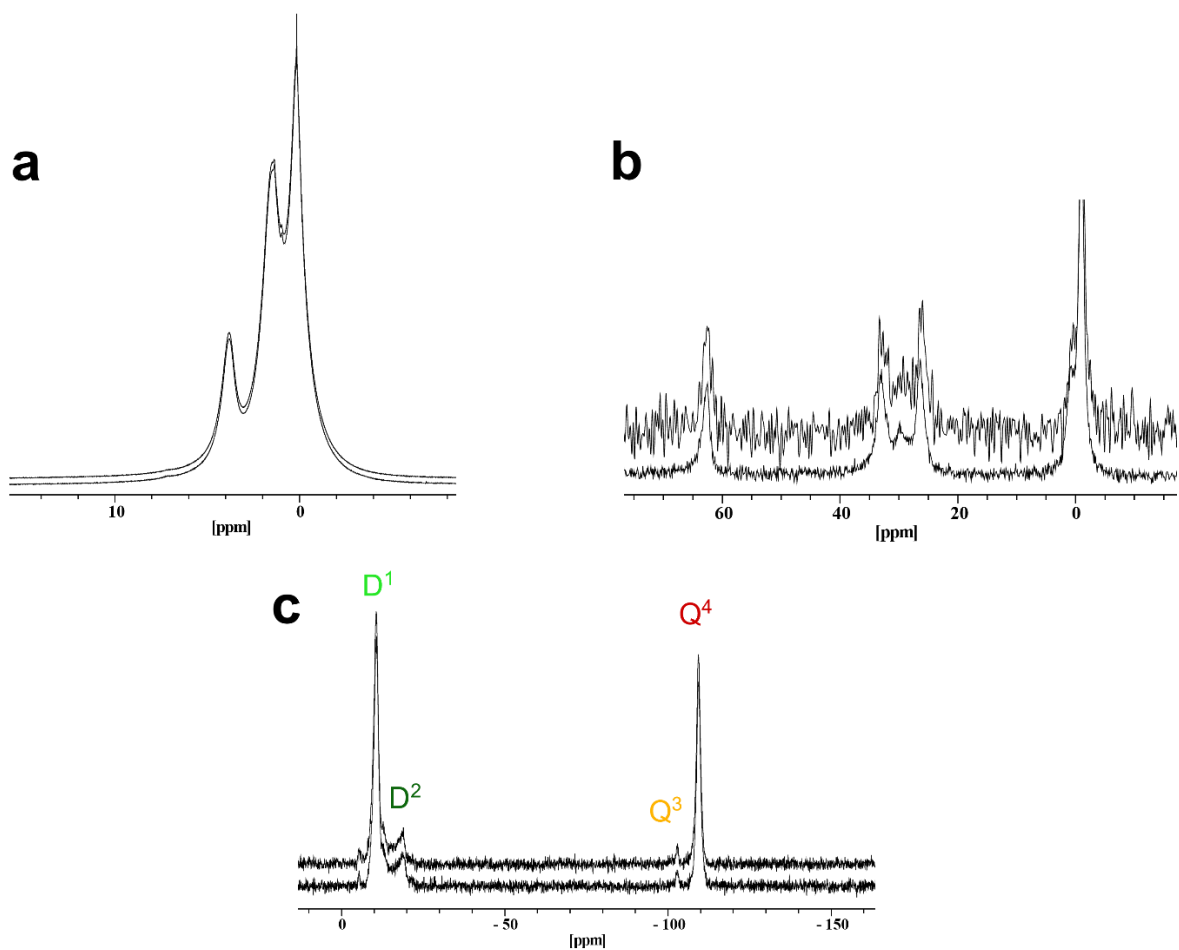

**Figure SI-18.** Investigating the stability of a PSiPOP material with 1,7-heptanediol linkers immersed in an aqueous solution of pH 3 (HCl). $^1$H (a); $^1$H → $^{13}$C CP MAS (b), and $^1$H decoupled $^{29}$Si single-pulse NMR spectra (c) of the parent material (bottom) and after immersed in pH 3 (top). All NMR spectra are almost identical before and after the immersion, proving the stability of this PSiPOP material.

## SI 2. Discussions

***SI 2.1. Self-polymerisation reaction of dichlorodimethylsilane with water resulting in $D^2$ Si species detected with solid-state $^{29}$Si MAS NMR:*** The $^1$H decoupled $^{29}$Si single-pulse MAS NMR spectra of PSiPOP materials reveal the presence of some $D^2$ coordinated Si atoms, ascribed to Si species in pure dimethylsilicone linkers (Figure 5f and table SI-1). The nature of such $D^2$ Si species originates from the self-polymerisation reaction of the dichlorosilane reagent with water. During the synthesis of PSiPOP materials, 2 sources of water can cause this polymerization: CySH crystal water and external water. First, the starting silicate D4R cubes are cyclosilicate hydrate (CySH) crystals which encage on average 5.33 water molecules per Si D4R, held together with a H-bonded network in a *LTA-like* topology.[S2] This crystal water is fixed in the small sod-like cage of the precursor. When the dichlorosilane is added to the (partially) dissolved silicate cubes, a condensation reaction of the CySH and the dichlorosilane occurs and the initial crystal structure of CySH is lost. The encaged water molecules are liberated from the small sod-like cage and could induce the self-polymerization reaction of the silane. A second source of water is called external water and could enter the reaction flask when adding different reagents. All possible precautionary measures were applied to limit the amount of external water: (i) sealing the reaction flask with high-vacuum grease and rubber stoppers, enabling a high vacuum (< 1mbar) (ii) septum sealed and/or high purity of all reagents and (iii) pre flushing every syringe and needle with N$_2$. Revealing a fraction of $D^2$ Si species with all these precautionary measures emphasizes the sensitivity of PSiPOP synthesis to any trace of water.

***SI 2.2. Absence of silanol resonances in $^1$H-$^1$H DQ-SQ MAS NMR :*** The absence of off-diagonal DQ correlations between the chemical shift region of silanol protons and the methyl protons of pentanediol in the $^1$H-$^1$H DQSQ MAS NMR spectra demonstrate the correct assignment of the D$^1$ Si species, i.e. linked to a diol. However, this D$^1$ Si resonances could also result from the presence of dimethylsilanol groups,[S3,S4] generated by the reaction of the octakis (dimethylsilyl chloride) spherosilicate with water. Absence of such silanol $^1$H species is obvious from demonstrating both the self-correlation of all $^1$H resonances and cross-correlation between the -CH$_3$ (of dimethylsilyl functionalities) and all -CH$_2$- protons (of pentanediol).

***SI 2.3. Unravelling the configuration of the reticular network of PSiPOP materials:*** The molar ratio of diol linkers compared to the numbers of Si-Cl in the spherosilicate solution during the synthesis is 1, which implies 2 possible configurations are possible (Figure SI-19): (i) an interconnected network of dimethylsilyl functionalized silicate cubes and diol linkers, where each diol linker is attached to 2 dimethylsilyl groups on both hydroxyl ends (further denoted as configuration 1) or (ii) a dendritic-like spherosilicate with terminal hydroxyl groups of the diol linker (denoted as configuration 2). The absence of alcohol functional groups in the 1D $^1$H spectra is a first hint of configurations. More evidence is provided by defining a $^1$H ratio, as follows:

$$^1H\ ratio = \frac{number\ of\ diol\ protons\ *\ \#\ diol\ linkers\ (D^1\ Si\ atoms)}{6\ *\ total\ \#\ of\ D\ Si\ atoms\ (D^1 + D^2\ Si\ atoms)}$$

Per cube, each Si atom in the D-region (D$^1$ & D$^2$) of the $^{29}$Si spectra is associated with a vast amount of protons, that is 6 (2x CH$_3$), independent of the configuration. 2 different methyl signals were fitted in the $^1$H spectra for their corresponding D$^1$ Si and D$^2$ Si signals (Figure SI-20), assuming different mobilities of the dimethyl silyl moieties depend on the linker (carbon, D$^1$ Si, or silicone, D$^2$ Si). On the contrary, the amount of protons of the diol linkers, related to the number of D$^1$ Si atoms, are dependent on the configuration and so a $^1$H ratio can be calculated for each configuration. A theoretical $^1$H ratio is calculated based on the $^{29}$Si NMR spectra, and for each resolved proton, O-**CH$_2$** and O-CH$_2$-(**CH$_2$**)$_n$ with n = 3 and 5 for 1,5-pentanediol and 1,7-heptanediol respectively, can be calculated. In case of ethylene glycol, 1 $^1$H signal of the carbon chain was measured, at 3.72 ppm, and so only 1 $^1$H ratio can be determined for both configurations (Figure SI-20a). For 1,5-pentanediol and 1,7-heptanediol each, 2 $^1$H signals were measured at 3.94 ppm (O-**CH$_2$**) and 1.59 ppm (O-CH$_2$-(**CH$_2$**)$_3$), and at 3.75 ppm (O-**CH$_2$**) and 1.74 ppm (O-CH$_2$-(**CH$_2$**)$_5$), respectively (Figure SI-20b and c). The calculated $^1$H ratios from the $^{29}$Si spectra are matched to those obtained after decomposition of the $^1$H spectra. In the case of 1,5-pentanediol, the calculation of the theoretical $^1$H ratios of the resolved protons is given below, for both configurations. The numbers of D$^1$ and D$^2$ Si atoms are given in table SI-1. For configuration 1, the theoretical $^1$H ratios are:

$$^1H\ ratio\ (\frac{O-CH_2}{CH_3}) = \frac{\#\ D^1\ Si\ atoms\ *\ 2}{\#\ (D^1 + D^2)\ Si\ atoms\ *\ 6} = \frac{6.6*2}{8.5*6} = 0.26$$

$$^1H\ ratio\ (\frac{O-CH_2-(CH_2)_3}{CH_3}) = \frac{\#\ D^1\ Si\ atoms\ *\ 3}{\#\ (D^1 + D^2)\ Si\ atoms\ *\ 6} = \frac{6.6*3}{8.5*6} = 0.39$$

For configuration 2, the theoretical $^1$H ratios are:

$$^1H\ ratio\ (\frac{O-CH_2}{CH_3}) = \frac{\#\ D^1\ Si\ atoms\ *\ 4}{\#\ (D^1 + D^2)\ Si\ atoms\ *\ 6} = \frac{6.6*4}{8.5*6} = 0.52$$

$$^1H\ ratio\ (\frac{O-CH_2-(CH_2)_3}{CH_3}) = \frac{\#\ D^1\ Si\ atoms\ *\ 6}{\#\ (D^1 + D^2)\ Si\ atoms\ *\ 6} = \frac{6.6*6}{8.5*6} = 0.78$$

All $^1$H ratios are summarised in table SI-2 and for all 3 diols, the $^1$H ratio from the $^1$H spectra agrees well with the theoretical calculations from the $^{29}$Si spectra for configuration 1. This confirms the chemical order of one diol molecule linking 2 corners of dimethylsilyl functionalized cube (Si$_8$O$_{20}$-Si(CH$_3$)$_2$-O(CH$_2$)$_n$O-Si(CH$_3$)$_2$-Si$_8$O$_{20}$).

**Table SI-2.** For each diol linker, their chemically resolved protons, and configuration, the $^1$H ratio is calculated according to the deconvolution of the 1D $^{29}$Si NMR spectra of Figure 5f. These ratios are compared to those retrieved from the deconvolution of the corresponding $^1$H NMR spectra (Figure SI-20). For each diol linker, the $^1$H ratios confirm the composition of configuration 1.

| Diol linker | Configuration | According to $^{29}$Si NMR $^1$H ratio | According to $^1$H NMR $^1$H ratio |
|---|---|---|---|
| ethylene glycol | | | |
| O-C**H**$_2$/C**H**$_3$ | 1 | **0.21** | 0.2 |
| | 2 | 0.42 | |
| 1,5-pentanediol | | | |
| O-C**H**$_2$/C**H**$_3$ | 1 | **0.26** | 0.26 |
| | 2 | 0.52 | |
| O-C**H**$_2$-(C**H**$_2$)$_3$/C**H**$_3$ | 1 | **0.39** | 0.39 |
| | 2 | 0.78 | |
| 1,7-heptanediol | | | |
| O-C**H**$_2$/C**H**$_3$ | 1 | **0.26** | 0.28 |
| | 2 | 0.52 | |
| O-C**H**$_2$-(C**H**$_2$)$_5$/C**H**$_3$ | 1 | **0.64** | 0.68 |
| | 2 | 1.28 | |

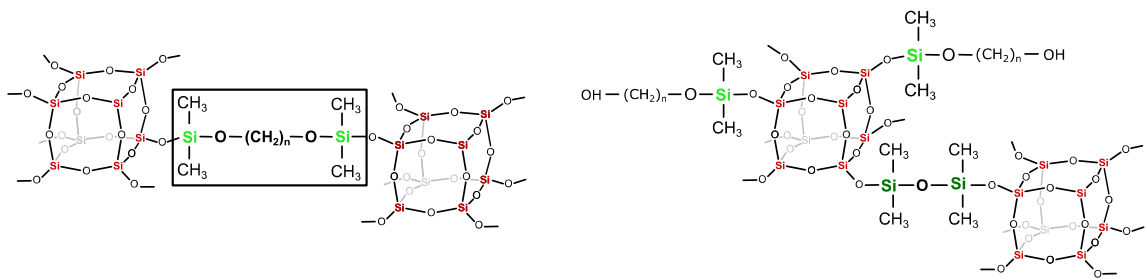

**Figure SI-19.** Representation of the chemical structure in which one diol linker reacted with both of its hydroxyl groups and is incorporated between 2 silicate cubes (configuration 1, left) and a dendritic like spherosilicate phase (configuration 2, right) with terminal hydroxyls. The D$^2$ Si fraction is assigned to the formation of some dimethylsilicone linkers, connecting some of the corners of neighboring silicate cubes.

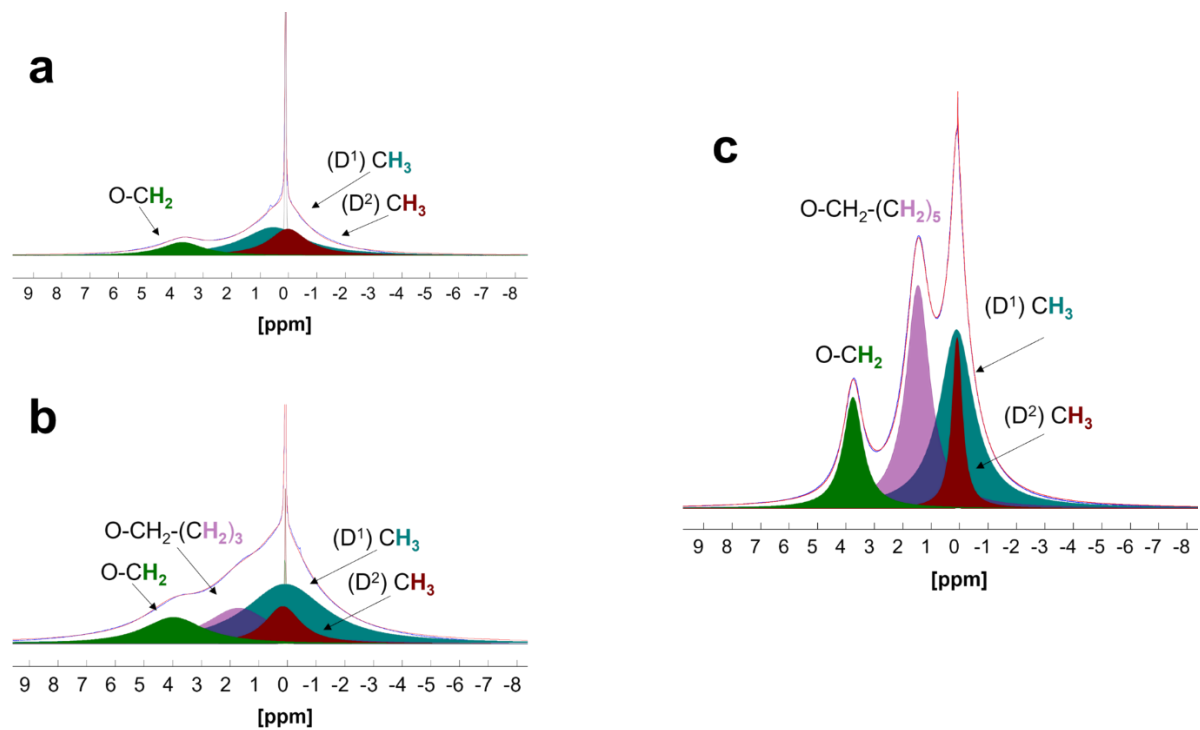

**Figure SI-20.** $^1$H spectra of ethylene glycol - (a), 1,5-pentanediol - (b), and 1,7-heptanediol (c) PSiPOP materials with spectra deconvolution. Individual, fitted signals are indicated in different colors.